\documentclass{applemlr}
\usepackage{amsmath}
\usepackage{enumerate} 
\usepackage{algorithm}
\usepackage{algpseudocode}
\usepackage{amsfonts}
\usepackage{amsthm}
\usepackage{newtxtt}
\usepackage{cleveref}
\usepackage{diagbox} 
\usepackage{colortbl}
\usepackage{amssymb}
\usepackage{xspace}
\usepackage{wrapfig}
\usepackage{adjustbox}
\usepackage{tabularx}
\usepackage{booktabs}
\usepackage{mathtools}
\usepackage{tikz}
\usepackage{enumitem}
\usepackage{silence}
\usepackage{dsfont}
\usepackage[table]{xcolor}
\usepackage[dvipsnames]{xcolor}
\usepackage{multirow}
\usepackage{makecell}
\usepackage{xfakebold}
\usepackage{tabularx}
\usepackage{graphicx}
\usepackage{xcolor}
\usepackage{multirow}
\usepackage{arydshln}
\usepackage{amsmath,amsfonts,bm}

\def\eqref#1{equation~\ref{#1}}
\def\1{\bm{1}}

\DeclareMathAlphabet{\mathsfit}{\encodingdefault}{\sfdefault}{m}{sl}
\SetMathAlphabet{\mathsfit}{bold}{\encodingdefault}{\sfdefault}{bx}{n}

\definecolor{textgray}{HTML}{6E6E73}
\usetikzlibrary{positioning, calc}
\usetikzlibrary{decorations.pathmorphing}

\makeatletter
\patchcmd{\wrong@fontshape}{\@gobbletwo}{}{}{}
\makeatother
\makeatletter
\AtBeginDocument{%
  \urlstyle{sf}
  
}
\makeatother

\numberwithin{equation}{section} 
\tcbuselibrary{minted}
\setminted[python]{
  linenos,
  breaklines,
  fontsize=\footnotesize,
  xleftmargin=2em
}

\definecolor{light}{RGB}{125, 125, 125}
\crefname{tcb@cnt@pbox}{code}{code}
\Crefname{tcb@cnt@pbox}{Code}{Code}
\crefname{assumption}{assumption}{assumption}
\Crefname{assumption}{Assumption}{Assumptions}

\newtcolorbox[auto counter]{pbox}[2][]{
  colback=white,
  title=Code~\thetcbcounter: #2,
  #1,fonttitle=\sffamily,
  fontupper=\sffamily,
  arc=2pt,
  colframe=bgcolor,
  coltitle=fgcolor,
  colbacktitle=bgcolor,
  toptitle=0.25cm,
  bottomtitle=0.125cm
}

\makeatletter
\newcommand\applefootnote[1]{%
  \begingroup
  \renewcommand\thefootnote{}%
  \renewcommand\@makefntext[1]{\noindent##1}%
  \footnote{#1}%
  \addtocounter{footnote}{-1}%
  \endgroup
}
\makeatother

\definecolor{cverbbg}{gray}{0.90}

\title{ 
{\color[HTML]{0000FF}{X}\color[HTML]{1A00E6}{-}\color[HTML]{3300CC}{T}\color[HTML]{4D00B3}{a}\color[HTML]{660099}{l}\color[HTML]{800080}{k}}: \\ On the Underestimated Potential of Modular Speech-to-Speech Dialogue System}

\author{
\parbox{\textwidth}{
Zhanxun Liu\textsuperscript{1,2,3}, Yifan Duan\textsuperscript{1,2}, Mengmeng Wang\textsuperscript{4}, Pengchao Feng\textsuperscript{1,2}, Haotian Zhang\textsuperscript{1}, Xiaoyu Xing\textsuperscript{1}, Yijia Shan\textsuperscript{1}, Haina Zhu\textsuperscript{1}, Yuxiang Zhao\textsuperscript{1}, Bei Liu\textsuperscript{5}, Yuhang Dai \textsuperscript{6}, Chaochao Lu\textsuperscript{3}, Xipeng Qiu\textsuperscript{2,7}, Lei Xie\textsuperscript{6}, Lan Wang\textsuperscript{8}, Nan Yan\textsuperscript{8}, Zilong Zheng\textsuperscript{4},\\ Ziyang Ma\textsuperscript{1}, Kai Yu\textsuperscript{1}, Xie Chen\textsuperscript{1,2,*}
}}

\affiliation{$^1$MoE Key Lab of Artificial Intelligence, X-LANCE Lab, Shanghai Jiao Tong University}
\affiliation{$^2$Shanghai Innovation Institute}
\affiliation{$^3$Shanghai AI Laboratory}
\affiliation{$^4$State Key Laboratory of General Artificial Intelligence, BIGAI}
\affiliation{$^5$The Hong Kong University of Science and Technology}
\affiliation{$^6$Audio, Speech and Language Processing Group, Northwestern Polytechnical University }
\affiliation{$^7$Fudan University }
\affiliation{$^8$Shenzhen Institutes of Advanced Technology, Chinese Academy of Sciences }

\contribution{\texttt{\{xcc\_zach,  chenxie95\}@sjtu.edu.cn} \\ \textsuperscript{*}Corresponding author.}
\abstract{
We present \textbf{X-Talk}, an open-source framework that champions a decoupled, modular design for LLM-driven speech-to-speech (S2S) systems. 
While the dominant trend favors end-to-end (E2E) modeling to optimize information flow, these "omni-models" often struggle to balance the competing objectives of complex speech tasks within a single network. 
X-Talk challenges this paradigm by demonstrating that a systematically optimized cascaded pipeline can achieve sub-second latency without sacrificing modular flexibility. 
Our framework seamlessly integrates specialized front-end components (e.g., VAD, speech enhancement) and diverse understanding models (e.g., ASR, emotion, and environmental sound analysis) with LLM capabilities like retrieval-augmented generation (RAG) and tool use. 
By revitalizing the cascaded approach, X-Talk highlights the underestimated potential of modular S2S systems and provides a robust foundation for future research and applications.
}
\metadata[Code]{\url{https://github.com/xcc-zach/xtalk}}
\metadata[Demo]{\url{https://xtalk.sjtuxlance.com}}
\date{\sffamily\today}

\begin{document}

\maketitle
% \begingroup
%   \renewcommand\thefootnote{} 
%   \footnotetext{\textsuperscript{*}Corresponding author.}
% \endgroup
\begin{figure*}[!ht]  
    \centering
    \includegraphics[width=0.98\linewidth]{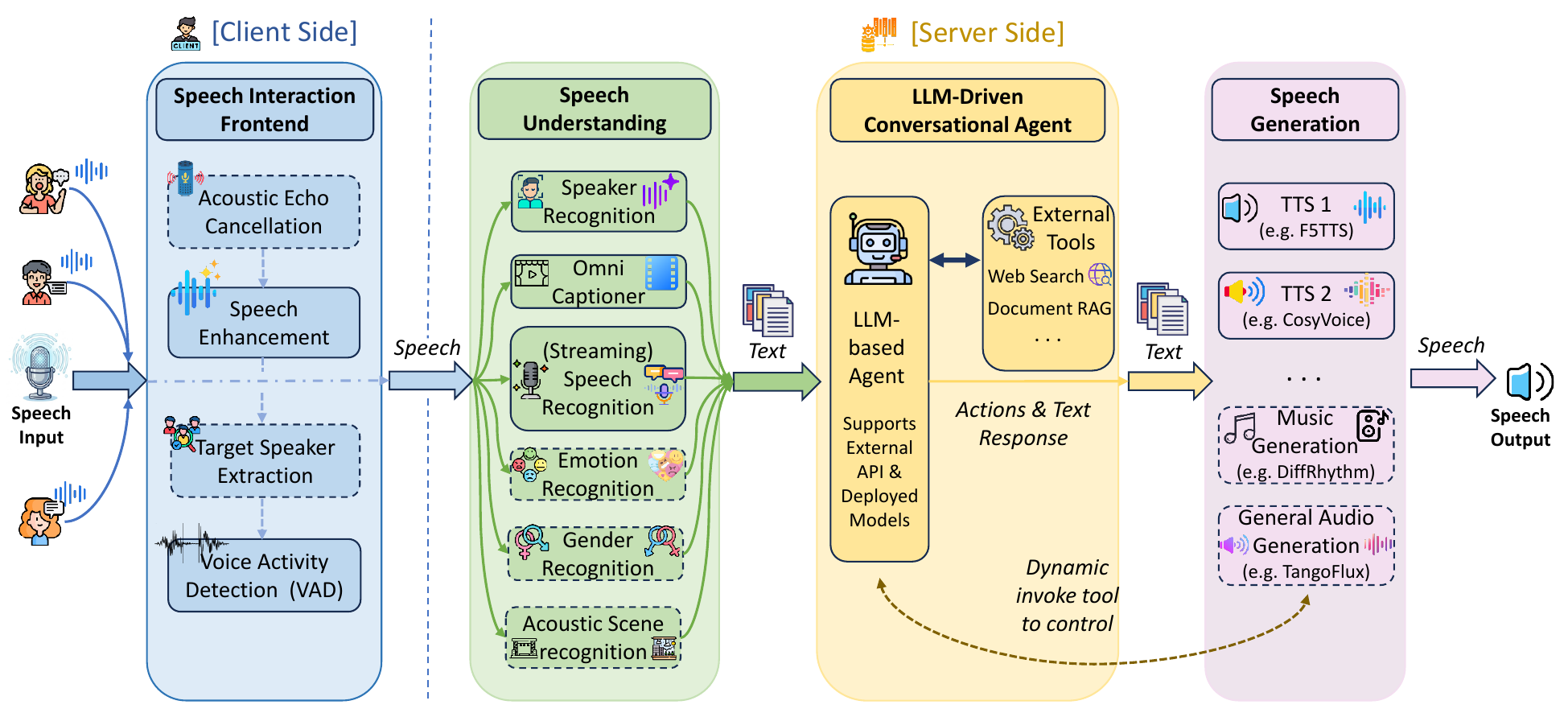}
    \caption{The data flow in the X-Talk framework, featuring four core functional modules of the speech-to-speech dialogue system: Speech Interaction Frontend, Speech Understanding, LLM-Driven Conversational Agent, and Speech Generation. 
    % The module indicated by the dashed box in the figure represents the component that X-Talk has not yet implemented but plan for future integration. 
    % In Speech Interaction Frontend, the noisy speech inputs first would undergo a series of front-end processing steps, such as echo cancellation and speech enhancement, and then enters the VAD module for speech detection. For Speech Understanding, the module uses various components such as speech recognition, emotion recognition to analyze and extract both linguistic and paralinguistic information embedded in speech. The rich information will be provided to the LLM to generate high-quality responses, with different tools for the LLM to invoke. For Speech Generation, different components are responsible for generating relevant and natural audio responses based on the outputs of the LLM. 
    The modules enclosed in solid lines denote currently integrated components, while those represented by dashed boxes indicate ongoing developments. 
}
    \label{fig:intro}
\end{figure*}

\section{Introduction}
\label{sec:intro}
% Speech-to-Speech (S2S) dialogue systems aim to facilitate fluent and effective human–machine interaction through natural spoken language, which are widely employed in AI-assisted applications~\citep{google, apple, amazon}. These systems can be broadly categorized into two distinct architectural paradigms: cascade systems and end-to-end systems. Recently, driven by advances in multi-modal large language models (MLLMs)~\citep{zhang-etal-2024-mm}, there has been a significant surge of interest in end-to-end approaches within the research and industrial communities. Nevertheless, despite claims that end-to-end systems offer low-latency responses, the ability to capture paralinguistic cues from user utterances, and seemingly unlimited potential ~\citep{xie2024mini}, they in fact face significant practical challenges that include high training costs, difficulty in modular optimization, and a tendency toward performance degradation, which is often colloquially referred to as "intelligence collapse". In this work, we revisit the cascaded dialogue system paradigm, aiming to challenge the prevailing perception that cascaded models are outdated and demonstrate that they can achieve performance on par with that of end-to-end approaches through careful architectural design and integration of various advanced speech models.
Speech-to-Speech (S2S) dialogue systems aim to facilitate fluent and effective human–computer interaction (HCI) through natural spoken language. 
These systems can be broadly categorized into two distinct architectural paradigms: 1) cascade systems and 2) end-to-end systems. 
Conventional cascaded systems utilize a pipeline composed of automatic speech recognition (ASR), natural language understanding (NLU), and speech-to-text (TTS) synthesis modules, which are widely employed in AI-assisted applications~\citep{google, apple, amazon}. 
Recently, driven by advances in multi-modal large language models (MLLMs)~\citep{zhang-etal-2024-mm}, there has been a significant surge of interest in end-to-end approaches within the research and industrial communities. 
Nevertheless, despite claims that end-to-end systems offer low-latency responses~\citep{xie2024mini, defossez2024moshi}, the ability to capture paralinguistic cues from user utterances~\citep{wang2024blsp, xue2024chat}, and the capacity to generate empathetic responses~\citep{wang2025opens2s, chen2025emova}, they face significant practical challenges. 
These include prohibitive training costs~\citep{zeng2024glm, defossez2024moshi}, a tendency toward intelligence  degradation~\citep{chen-etal-2025-slam}, and inherent unreliability~\citep{ma2025towards, yang2025speech}, which often renders them unsuitable for deployment in real-world user environments.
In this work, we revisit the cascaded dialogue system paradigm, aiming to challenge the prevailing perception that cascaded models are outdated. 
\textbf{We demonstrate that through meticulous architectural design and the integration of advanced speech models, cascaded systems can achieve performance on par with, or even exceeding, that of end-to-end approaches.}

% As shown in Figure~\ref{fig:intro}, we begin by systematically decomposing the speech-to-speech dialogue system into four core functional modules:
To operationalize this vision, a highly modularized framework is designed to overcome the inherent limitations of current monolithic models. Specifically, as illustrated in Figure~\ref{fig:intro}, we systematically decompose the speech-to-speech dialogue process into \textbf{4 Core Functional Modules}:

\begin{itemize}
\item[(1)] \textbf{Speech Interaction Frontend}: This module processes noisy speech inputs through a set of preprocessing components, including \textit{speech enhancement}, \textit{voice activity detection (VAD)}, and \textit{target speaker extraction}, which are designed to be lightweight and operate in a streaming manner. In this work, the frontend models are deployed on the client side to minimize end-to-end latency.

\item[(2)] \textbf{Speech Understanding}: This module aims to analyze and extract both linguistic and paralinguistic information embedded in speech. It incorporates components such as \textit{speech recognition}, \textit{speaker identification}, \textit{gender} and \textit{emotion recognition}, and \textit{acoustic scene recognition}. In addition, detailed audio captioning models can be employed to achieve a more comprehensive understanding of the speech signal and to convert relevant information into textual representations.

\item[(3)] \textbf{LLM-Driven Conversational Agent}: This module consumes textual inputs produced by the ASR system and other upstream components of the speech understanding module, and performs high-level dialogue reasoning and response generation. \textit{Large Language Models (LLMs)} serve as the core engine, augmented with auxiliary components such as a \textit{web searcher}, a \textit{local database retriever}, and a \textit{reasoning (thinking) module} to handle complex queries and to integrate information from multiple heterogeneous sources.

\item[(4)] \textbf{Speech Generation}: This module is responsible for generating relevant and natural audio responses based on the outputs of the LLM. Various \textit{speech, audio, and music generation models} can be employed as generation agents. Under this design, multiple speech generation models can be coordinated and organized as a chorus, enabling flexible and expressive audio output.

\end{itemize}

This modular decomposition enables fine-grained control, targeted optimization, and seamless integration of linguistic, paralinguistic, and environmental signals, laying the foundation for a highly capable, interpretable, and reliable dialogue system. 

Then, we propose the \textbf{Event-driven Architecture} to orchestrate the entire cascaded system in a novel and principled manner. Specifically, inputs and outputs of all modules are uniformly encoded as structured events and propagated along a central event bus. This design not only enhances modularity and interoperability but also facilitates asynchronous and dynamic coordination among components. Furthermore, to reduce system response latency and improve the naturalness of interaction, we refine the turn-taking logic in the VAD module and redesign relevant modules to operate in a streaming fashion. We also introduce a buffered statement mechanism, which generates buffered response content ahead of the tool call to mask the latency introduced by external tool calls. 

We make and integrate the above design into \textbf{X-Talk}, an open-source spoken dialogue system framework tailored for real-world deployment. X-Talk not only enables low-latency, paralinguistic-aware, full-duplex spoken interaction but also offers high modularity: individual components can be independently maintained, retrained, or optimized, and new modules can be seamlessly integrated without retraining the overall pipeline. This architecture thus achieves a compelling balance between high performance and fine-grained controllability, making it well-suited for practical, scalable, and evolving conversational AI applications. 

In summary, X-Talk provides the following features that distinguish it from other models or frameworks: 
\begin{itemize}
    \item \textbf{A low-latency, full-duplex dialogue system} that significantly reduces interaction delay by optimizing audio stream processing and response scheduling mechanisms, enabling more natural and fluid human–computer interaction.
    \item \textbf{A modular, event-driven architecture} that coordinates heterogeneous functional components. This design not only ensures flexible integration and maintainability but also unlocks the latent paralinguistic understanding capabilities inherent in cascaded spoken dialogue systems.
    \item \textbf{A fully open-source framework, featuring strong engineering merits} with complete model controllability, a lightweight asynchronous server implementation, and native support for high-concurrency deployment, collectively enabling robust and scalable real-world productization.
\end{itemize}

\section{Related Work}
\label{sec:related work}
\subsection*{Cascaded S2S Dialogue System}
Cascaded spoken dialogue systems have been a central paradigm in human–computer interaction. Early systems, such as the \textit{Let’s Go Public spoken dialog system}~\citep{raux2005lets}, demonstrated the feasibility of robust deployment in real-world scenarios. Recent integration of LLMs has substantially enhanced cascaded architectures~\citep{chen2025fireredchatpluggablefullduplexvoice, arora-etal-2025-espnet, xiaozhi_esp32}. Nevertheless, most contemporary cascaded dialogue systems still follow a rigid three-stage pipeline: automatic speech recognition (ASR), LLM-based language understanding and generation, and text-to-speech synthesis (TTS). This modular design limits the exploitation of paralinguistic cues (e.g., intonation, emphasis, emotional prosody), which are essential for natural and context-aware interaction~\citep{schuller2011speaker}. Furthermore, many open-source and commercial frameworks, such as TEN~\citep{tenai}, Voicechat2~\citep{voicechat2}, Vocal-Agent~\citep{vocal-agent}, Bolna~\citep{bolna} RealtimeVoiceChat~\citep{RealtimeVoiceChat}, LiveKit Agent~\citep{LiveKit}, Pipecat~\citep{Pipecat} depend on cloud-based APIs for core components, raising privacy and security concerns.

\subsection*{End-to-End S2S Dialogue System}
End-to-end spoken dialogue systems aim to unify speech perception, language understanding, dialogue management, and speech generation within a single neural architecture. Recent advances, exemplified by Moshi~\citep{defossez2024moshi}, Mini-Omni~\citep{xie2024mini}, GLM-4-Voice~\citep{zeng2024glm}, GPT-4o~\citep{hurst2024gpt}, SLAM-Omni~\citep{chen-etal-2025-slam}, Kimi Audio~\citep{ding2025kimi}, MIMO-Audio~\citep{coreteam2025mimoaudio}, Moss-Speech~\citep{MOSS-Speech}, SALMONN-omni~\citep{yu2025salmonn}, Qwen2.5-Omni~\citep{Qwen2.5-omni} and Qwen3-Omni~\citep{Qwen3-Omni} , highlight the growing maturity of this paradigm.
Current systems largely focus on mitigating performance loss when transitioning from text-based to spoken modalities. Despite their architectural elegance, they still underexploit paralinguistic cues~\citep{chen2025audiollmsreallylisten}, which are vital for natural interaction. Moreover, the integration of advanced capabilities such as retrieval-augmented generation (RAG) and tool/function calling in fully differentiable speech-grounded dialogue remains at a preliminary stage~\citep{chen-etal-2025-wavrag, feng-etal-2025-enhancing}. While such functionalities are well-established in text-based LLMs, adapting them to end-to-end speech systems raises challenges in latency from web searching and reliability.
\section{Design of X-Talk}
\label{sec:design}
\begin{figure*}[!ht]  
    \centering
    \includegraphics[width=1\linewidth]{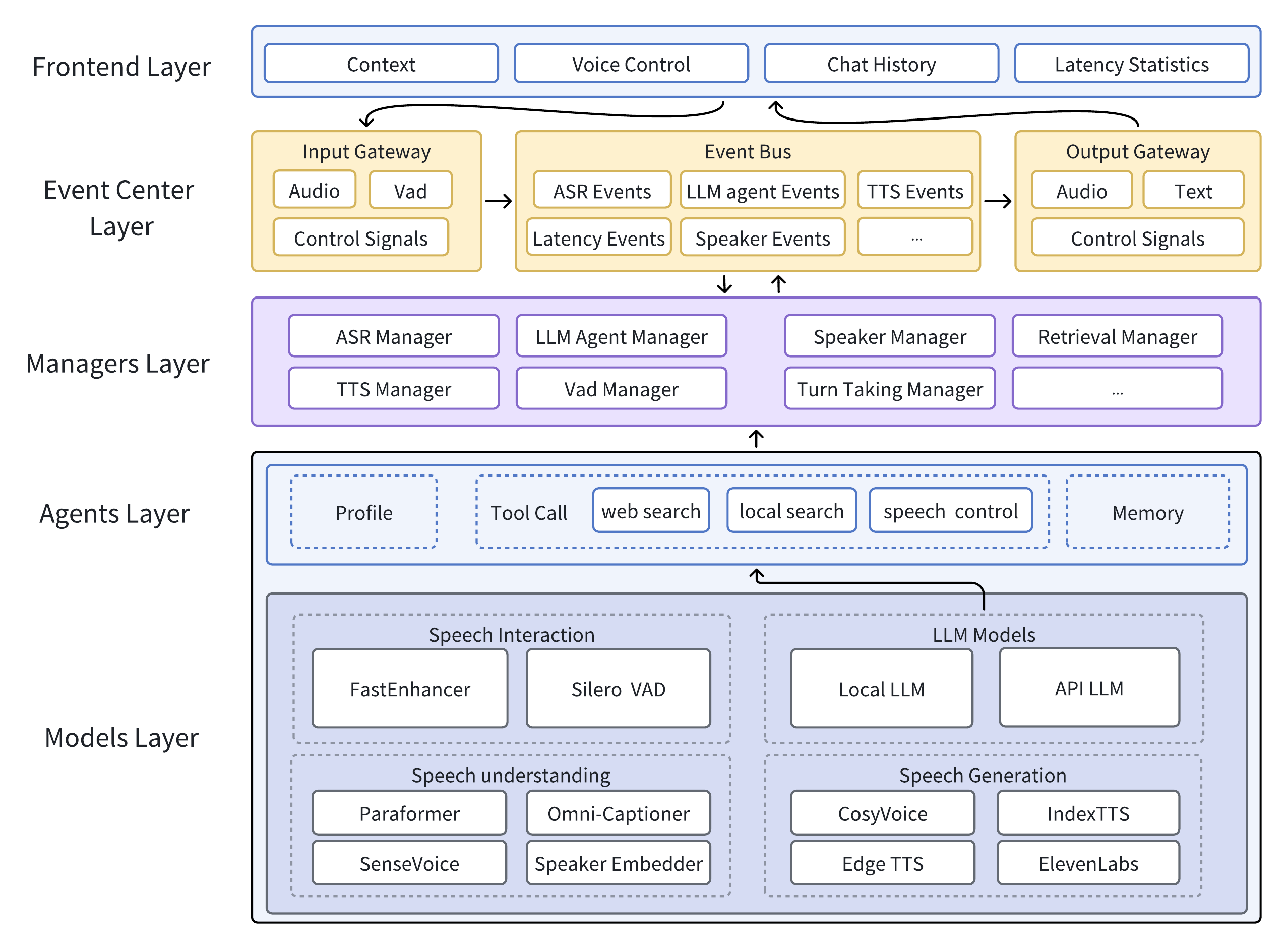}
    \caption{The system architecture in X-Talk. 
    The entire system is built around a centralized event bus with a hierarchical design. All layers communicate asynchronously through an event publish-subscribe mechanism, enabling efficient management of complex conversational state and data flow. 
    % The Frontend Layer serves as the user-facing interface and directly handles client interaction. The Event Center Layer acts as the system’s communication hub, unifying event routing and protocol translation. The Managers Layer orchestrates the core conversational workflow through specialized managers. The Agents Layer functions as the system’s task-planning and execution engine, integrating structured inputs from upstream models into a coherent speech and audio understanding. The Models Layer provides a unified, interface-driven abstraction for core speech-to-speech dialogue capabilities.
    }
    \label{fig:architecture}
\end{figure*}
\subsection{Overall Architecture of X-Talk}

% The system follows a modular, stage-wise functional flow—progressing from noisy speech input, through frontend speech interaction, speech understanding, and LLM-driven conversational agent, to speech generation. This logical pipeline is implemented via a layered, event-driven, and loosely-coupled architecture, which forms the core of X-Talk. This design systematically addresses the key challenges of real-time speech-to-speech dialogue systems: controlling sub-second level latency, orchestrating multiple components, and enabling flexible integration and swapping of backend models and services. As shown in Figure~\ref{fig:architecture}, the entire system is built around a centralized event bus, with each layer communicating asynchronously via event publishing/subscribing, enabling efficient management of complex conversational state and data flow.

The architecture of X-Talk follows a modular, stage-wise functional pipeline, transitioning from raw acoustic signals through frontend interaction, speech understanding, and LLM-driven reasoning and integration, to final speech generation. 
As illustrated in Figure~\ref{fig:architecture}, this architecture is realized through a layered, event-driven, and loosely-coupled architecture. 
The system is anchored by a centralized event bus, where each layer communicates asynchronously via a publish-subscribe mechanism, enabling efficient management of complex conversational states and data flow. 
This design is specifically engineered to address the core challenges of real-time speech-to-speech dialogue systems: achieving sub-second latency, orchestrating multiple components, and ensuring the flexible integration of backend models and services. 

\paragraph{Frontend Layer} This layer acts as the user-facing interface, directly handling frontend interaction. It provides the conversational UI, performs client-side Voice Activity Detection (VAD), audio denoise and enhancement, and displays real-time latency metrics to the user. It packages audio streams, VAD markers, and context for transmission to the backend.

\paragraph{Event Center Layer} This layer serves as the system's communication hub, unifying event routing and protocol translation. It comprises three components: 1) the Input Gateway converts frontend streams into typed events; 2) the Output Gateway delivers processed events back to the frontend; and 3) the Event Bus provides the asynchronous messaging fabric, routing events between all components.
Together, they decouple all other layers, handling protocol adaptation, event distribution, and lifecycle isolation—forming the extensible backbone of the architecture.

\paragraph{Managers Layer} This layer orchestrates the core conversational workflow through specialized, capability-specific managers. Each manager is dedicated to a particular component and operates by subscribing to relevant events, executing its logic, and publishing new events to drive the dialogue forward.

\paragraph{Agents Layer} This layer serves as the system’s task-planning and execution engine, integrating structured inputs from upstream models and historical context to form a coherent speech understanding. The agent then orchestrates tool use, including web search, local retrieval, audio control, to collect needed data or perform actions. Finally, it synthesizes the retrieved or processed information into a context-aware, coherent natural language response.

\paragraph{Models Layer} This layer provides a unified, interface-driven abstraction for core speech-to-speech dialogue system capabilities: speech understanding, LLM agent, and speech generation. It defines stable, modular contracts for each capability, enabling compliant implementations to be seamlessly integrated, swapped, or scaled without impacting other system components. 

\subsection{Design Details in X-Talk}

\subsubsection{Speech Interaction Frontend}
% In Speech Interaction Frontend, the noisy speech inputs will be processed by different reprocessing components, such as speech enhancement and VAD. Currently, X-Talk deploys these models on the client side to reduce the latency. 
%  X-Talk should be robust in noisy and acoustically complex environments. 
In real-world deployments, background noise can easily influence the speech input stream, which may lead to unstable turn-taking and influence backend interactions. 
Meanwhile, a good interaction experience requires the system to respond to user's speech and barge-in with low latency. 
% Therefore, X-Talk introduces a streaming speech enhancement model and a VAD model on the client. 
X-Talk addresses these requirements by deploying a streaming speech enhancement model and a VAD module directly on the client side. 
% X-Talk first enhances the speech inputs using FastEnhancer~\citep{ahn2025fastenhancerspeedoptimizedstreamingneural} via streaming, and the system is going to allow for switching to alternative models. The enhanced speech is then provided to the VAD to reduce the likelihood of false detections and false interruptions. Specifically, X-Talk currently uses Silero VAD to analyze speech frames in real time~\citep{SileroVAD}. Upon detecting human speech, the client pauses playback immediately and sends a ``vad\_speech\_start'' event to the server for notification. In this way, X-Talk supports real-time barge-in. Users can speak while the system is talking, and the system can be interrupted with low latency.
Specifically, X-Talk employs FastEnhancer~\citep{ahn2025fastenhancerspeedoptimizedstreamingneural} to perform streaming denoising on raw audio inputs. This enhanced signal is then fed into Silero VAD~\citep{SileroVAD} to analyze speech frames in real time, significantly reducing false triggers in acoustically complex environments. 
Upon detecting human speech, the client immediately pauses local playback and dispatches an event to the server. This client-side execution minimizes the round-trip latency for interruptions, allowing users to naturally interrupt the system while it is speaking. 

% To improve X-Talk’s robustness in noisy and complex environments, we deploy a speech enhancement model to process the speech inputs in real time via streaming.  X-Talk currently employs FastEnhancer as the speech enhancement model~\citep{ahn2025fastenhancerspeedoptimizedstreamingneural}, and the system will allow for switching to alternative models later. The enhanced speech is then fed into the VAD to reduce the likelihood of false detections and false interruptions.
% To reduce latency and ensure accurate speech detection, X-Talk deploys the VAD model directly on the client side to analyze users' speech input in real-time. Specifically, X-Talk currently uses Silero VAD to analyze speech frames in real time on the frontend~\citep{SileroVAD}.
% When human speech is detected, the system immediately pauses the current playback to achieve low-latency interruptions and sends a ``vad\_speech\_start'' event to the server for notification. In this way, X-Talk supports real-time barge-in. Users can speak while the system is talking, and the system can be interrupted with low latency.

\subsubsection{Speech Understanding}
% In many cascaded spoken dialogue systems, the system typically relies solely on ASR to transcribe user speech into text and pass it to downstream models, thereby overlooking the rich paralinguistic information inherent in speech (e.g., background noise, the speaker’s emotional state). This loss of information makes it difficult for the system to fully exploit the original speech signal, which in turn limits the depth of understanding and interactive capabilities of the dialog system. Therefore, we aim to introduce a broader set of models into the speech understanding pipeline to extract richer paralinguistic information for the LLM, thereby improving X-Talk’s ability to comprehend the speech.
Conventional cascaded systems often rely solely on ASR to transcribe speech to text, thereby discarding rich paralinguistic cues such as emotional state and acoustic context. 
This information loss limits the system’s depth of understanding. 
To mitigate this, X-Talk integrates a multi-dimensional understanding pipeline that extracts linguistic, paralinguistic, and environmental information.

\paragraph{Speech Recognition}
X-Talk supports multiple ASR backends, including Zipformer~\citep{yao2023zipformer}, SenseVoice~\citep{an2024funaudiollmvoiceunderstandinggeneration}, and Paraformer~\citep{paraformer}, covering both streaming and offline scenarios. 
To balance between transcription accuracy and real-time requirements, we implement a pseudo-streaming mechanism for offline models. 
We maintain a cumulative buffer and an incremental buffer. For each new audio chunk, the system performs recognition on the cumulative buffer and appends the result to a stable prefix. We use a sliding-window cache to monitor the stability of the generated text; once the prefix remains unchanged for a predefined window size, it is finalized, and the corresponding audio is flushed from the buffer. This approach delivers a streaming-like user experience while leveraging the superior accuracy of offline models.

\paragraph{Acoustic Context Awareness}
% In X-Talk, We employ Qwen3-Omni-30B-A3B-Captioner~\citep{Omni-Captioner}, a powerful audio analysis model that generates accurate and comprehensive descriptions of user's audio scenes and conveys these descriptions to the large language model (LLM). With this method, X-Talk gets the ability to perceive the acoustic environment, partially compensating for the loss of paralinguistic information.
% Specifically, X-Talk maintains a 15-second rolling audio buffer to continuously receive unprocessed raw speech. X-Talk invokes Qwen3-Omni-Captioner at fixed intervals to generate scene descriptions based on the speech currently in the buffer. Recognizing that the captioner’s output may occasionally be verbose or lack focus, we provide an optional caption-rewriter module. It uses an LLM to rewrite the scene description to make it more concise and precise.
% During subsequent conversations, when the system needs to use the LLM to generate a response, it retrieves the latest speech scene caption and inserts it into the LLM's system prompt. Through this approach, the LLM can continuously perceive dynamic speech scenarios and generate more intelligent responses that align with the user's situation.
To perceive the user's environment, we integrate Omni-Captioner~\citep{Omni-Captioner}, a powerful audio analysis model that generates detailed descriptions of audio scenes. 
We maintain a 15-second rolling audio buffer and invoke the captioner at fixed intervals. An optional LLM-based caption-rewriter module is employed to condense verbose outputs into precise, context-rich summaries. These captions are dynamically injected into the LLM’s system prompt, enabling the agent to generate responses that are contextually aligned with the user's physical surroundings. 

\paragraph{Speaker Recognition}
% To enable X-Talk with the ability to perceive the identity of the speaker, we have integrated a speaker recognition model. Currently, we have configured WeSpeaker ResNet34 based on the pyannote.audio framework~\citep{Wang2023,bredin23_interspeech} and it is extensible to support other models. In each turn of conversation, the speaker recognition model extracts voiceprint features from the user's speech input and matches them against a registered speaker voiceprint database to identify the current speaker in real time. Specifically, during the initial interaction, the system registers the user as Speaker 1. In subsequent turns, the system compares the current voiceprint with the voiceprint embeddings of known speakers. If the similarity exceeds a predefined threshold, the speaker is identified accordingly. Then, the system dynamically updates the speaker's voiceprint embedding using exponential moving average to enhance the robustness of identity recognition over extended conversations. Or if the current voiceprint doesn't match any registered speaker in the registered speaker voiceprint database, the system will register the user as a new speaker. After identifying the speaker's identity, the system can insert the identity into the prompt and input it into the LLM. Through this mechanism, X-Talk effectively distinguishes between different speakers in multi-user dialog scenarios.
Identity awareness is facilitated through a speaker recognition module based on the Pyannote framework using WeSpeaker ResNet34~\citep{Wang2023,bredin23_interspeech}. 
During the initial interaction, the system registers the user’s voiceprint. In subsequent turns, the module extracts embeddings from the input stream and performs real-time matching against a registered database. We utilize an Exponential Moving Average (EMA) update strategy to adapt the stored embeddings over time, ensuring robustness against intra-speaker variability. The identified speaker ID is passed to the LLM, enabling personalized multi-user interaction. 

\subsubsection{LLM-Driven Conversational Agent}

% In cascaded speech-to-speech dialog systems, while LLMs excel at text comprehension and generation, they remain constrained by limitations such as the inability to access external information in real time or dynamically control system states. To further enhance the LLM's capabilities and deliver a more natural and feature-rich interactive experience, X-Talk uses LangChain\footnote{\url{https://github.com/langchain-ai/langchain}} to configure a suite of tools for the LLM to invoke. This empowers the LLM with retrieval and internal system control capabilities. Furthermore, X-Talk's flexible architecture supports the continuous addition of new tools to expand the system's capability boundaries.
While LLMs excel at reasoning, they are inherently limited by their training knowledge cutoff and lack of system state dynamic control. 
X-Talk leverages LangChain\footnote{\url{https://github.com/langchain-ai/langchain}} to equip the LLM with a specialized toolset, transforming it into an autonomous agent capable of real-time information retrieval and system state management.

% \subsubsection{Retrieve Tools}
\paragraph{Web Search} 
% We have implemented a web search tool for the LLM, enabling it to retrieve external information in real time and provide users with a more timely and accurate communication experience. 
% % The tool performs web searches by initiating HTTP requests via the Serper API\footnote{\url{https://serper.dev/}} and formats the results into text. 
To provide users with up-to-date and factually accurate information, we have implemented a Web Search tool that enables the LLM to access external data in real time. 
The tool executes queries via the Serper API\footnote{\url{https://serper.dev/}} and processes the retrieved content into a structured format. 
% Building upon this foundation, we have added a retrieval content refinement feature to prevent the model from receiving incorrect or irrelevant information. First, we compute semantic coverage using word embeddings (from Qwen2.5-0.5B-Instruct~\citep{Yang2024Qwen25TR}).
% Based on this coverage score, we apply a tiered processing strategy. 
To prevent the model from ingesting irrelevant or erroneous information, we plan to employ a tiered semantic filtering strategy based on word embeddings generated by Qwen2.5-0.5B-Instruct~\citep{Yang2024Qwen25TR}. 
For results with high coverage, we directly use their snippets. For results with middle coverage, we further fetch the full content of the linked page and provide it to the LLM. For results with low coverage, we discard them entirely. 
% We register this tool within the LLM and enable it to proactively invoke the tool whenever it determines that its response requires real-time information or needs to reference external web content.
The LLM proactively invokes this tool whenever it identifies a need for contemporary data or external verification.

\paragraph{Local Search} 
% To further optimize personalized user experiences, we have implemented local knowledge base construction and retrieval capabilities. X-Talk supports users uploading files in multiple formats to build session-independent local vector databases. 
To enhance personalized user experiences, X-Talk supports RAG through a local knowledge base. 
Users can upload heterogeneous file formats to construct session-independent vector databases. 
Specifically, X-Talk uses Qwen3-Embedding-0.6B as the embedding model and uses the Chroma vector database for the storage and management of document vectors~\citep{qwen3embedding}. 
% We have equipped the LLM with efficient local retrieval tools, enabling it to access user file content in real time and provide more accurate responses. The LLM will proactively invoke this tool when it thinks that its response requires information from the user's document.
These tools allow the LLM to perform precise, real-time retrieval of user-specific documents, ensuring that responses are grounded in the user's personal data or specialized domain knowledge.

% \subsubsection{Speech Control Tools}
\paragraph{Timbre Switching} 
% During system interaction, we aim for X-Talk to automatically switch to the user's desired timbre based on their needs. To achieve this, we have equipped the LLM with a timbre-switching tool. 
% Through the LLM's outstanding text understanding capabilities, the LLM can identify the user's timbre preferences and autonomously invoke the tool to control the system's voice. When the timbre switching tool is invoked, the system automatically maps the timbre to its corresponding reference audio and switches the reference audio for the TTS, thereby achieving the timbre switching functionality.
A key feature of X-Talk is its ability to autonomously adapt its vocal persona based on user requirements. 
We have equipped the LLM with a Timbre Switching tool, leveraging its advanced natural language understanding to identify explicit or implicit user preferences for specific voices. Upon invocation, the system dynamically maps the desired timbre to corresponding reference audio and updates the TTS engine’s configuration on-the-fly, achieving seamless transitions between diverse vocal identities.

\paragraph{Emotion Switching} 
% In cascaded dialog systems, the system should be able to automatically adjust its speech emotional output based on the semantic content expressed by the user. When responding, it should respond with natural and varied emotional expression to make interactions more authentic. We rely on the powerful text comprehension capabilities of LLMs to achieve flexible switching of emotional states. To support different types of TTS models, we currently implement two emotion switching strategies. For models such as IndexTTS~\citep{deng2025indextts}, which do not natively support direct emotion control, we switch among pre-defined emotional reference audio samples within the same voice profile to achieve emotional variation. For models such as IndexTTS2~\citep{zhou2025indextts2}, which provide native emotion control, we directly modify the emotion vector to produce speech with the desired affect.
To foster authentic human-machine interaction, the system should be able to adjust its emotional prosody based on the semantic content expressed by the user. 
We implement a flexible emotion switching strategy that accommodates different TTS architectures. For models without native emotion control, such as IndexTTS~\citep{deng2025indextts}, the system switches between pre-defined emotional reference audio prompt within the same voice profile. 
For models supporting native emotion control, such as IndexTTS~\citep{deng2025indextts}, the LLM directly modifies the emotion vectors. 
The implementation of emotion switching is decoupled from the system’s core logic and handled independently by each TTS class, providing a flexible and extensible foundation for integrating additional TTS models in the future.

\paragraph{Adaptive Thinking}
% In spoken dialogue systems, enabling large language models to “think” while interacting in real time is highly challenging, as it involves issues such as system latency and response coherence. Recent studies on end-to-end dialogue systems have made attempts to augment speech-language models (SLMs) with deliberation capabilities~\citep{chiang2025stitchsimultaneousthinkingtalking, chiang2025shankssimultaneoushearingthinking, shih2025can, xie2025mini}. However, these approaches generally suffer from poor controllability or lack flexibility in determining when deliberation is necessary.
Enabling thinking within real-time spoken dialogue systems poses significant challenges regarding latency and coherence. 
Unlike recent end-to-end approaches that attempt simultaneous "thinking" and "talking" through a unified modeling~\citep{chiang2025stitchsimultaneousthinkingtalking, xie2025mini}, which often suffer from limited controllability and reliability, X-Talk treats thinking as an explicit, event-driven state.
In X-talk, we design thinking as an explicit event handled by a dedicated Thinking module. When the dialogue system requires deliberation, it can invoke this event on demand, offering high flexibility and controllability without introducing additional latency to normal interactions.

\subsubsection{Speech Generation}
% In Speech Generation module, X-Talk aims to combine multiple models(e.g.,TTS, Music Generation) and has them work collaboratively to produce flexible and expressive audio output based on the outputs of the LLM. At present, X-Talk has only integrated TTS, and integration of other models is currently underway.

% \paragraph{Text to Speech}
% We have deployed a range of high-quality TTS models(e.g.,IndexTTS~\citep{deng2025indextts,zhou2025indextts2}, CosyVoice~\citep{du2024cosyvoice}) in X-Talk to convert model responses into speech. In addition, we have adopted vLLM in parts of model deployments to boost inference throughput and speed up response generation~\citep{kwon2023efficient}. At the same time, X-Talk’s architecture makes it easy to integrate additional TTS models. Beyond that, within X-Talk, we are trying to regard the TTS model as a new tool and enable the LLM to control it through tool calls. Currently, X-Talk supports using LLMs to control the timbre and emotion of the speech generated by the TTS model, and we are trying to further explore potential of this approach.

The Speech Generation module is designed as a collaborative ``chorus'' of diverse synthesis models such as speech synthesis and music generation. Currently, X-Talk integrates high-fidelity TTS models such as CosyVoice~\citep{du2024cosyvoice} and IndexTTS~\citep{deng2025indextts,zhou2025indextts2}, and integration of other models is underway. 
To maximize inference throughput, we employ vLLM~\citep{kwon2023efficient} for model deployment. 
By treating the TTS engine as a controllable tool, the LLM can adjust prosody and emotional expression on-the-fly, achieving a level of expressiveness that rivals end-to-end systems while maintaining the precision from the cascaded designs. 
\section{Modularity and Extensibility}
\subsection{Supported Models}
Table~\ref{tab:integrated_modles} presents the models integrated in X-Talk, which are organized into four functional modules covering the full speech interaction pipeline. 
% Each module is responsible for different types of information and capabilities required in spoken language interaction, ranging from signal-level processing to high-level reasoning and controllable speech generation.
These components manage the full spectrum of spoken dialogue, from low-level signal processing to high-level cognitive reasoning and expressive synthesis.
The system also incorporates a mix of state-of-the-art open-source models and industry-leading proprietary APIs, demonstrating the framework's broad compatibility. 

% The frontend and speech understanding modules capture both linguistic content and paralinguistic information from speech, enabling richer and more context-aware representations. Built on top of these inputs, the language model module maintains strong reasoning and task-solving capabilities, without being weakened by the integration of multimodal information. Finally, the speech generation module provides fine-grained control over synthesized speech, allowing X-Talk to produce expressive and adaptive spoken responses.
\

\definecolor{groupgray}{RGB}{245,245,245}
\newcolumntype{L}{>{\raggedright\arraybackslash}p{4cm}}  
\newcolumntype{M}{>{\raggedright\arraybackslash}p{8cm}} 
\newcolumntype{R}{>{\centering\arraybackslash}p{4cm}}   

\begin{table}[htbp]
\centering
%\small  % 缩小字体

\begin{adjustbox}{max width=0.95\linewidth}  % 限制最大宽度
\begin{tabular}{@{}LMR@{}}
\toprule

\rowcolor{groupgray}
% \multicolumn{3}{@{}l}{\textbf{Module Category}} \\
\textbf{Component} & \textbf{Model} & \textbf{Deployment} \\
\midrule

\rowcolor{groupgray}
\multicolumn{3}{@{}l}{\textbf{Speech Interaction Frontend}} \\
Enhancement   & FastEnhancer                  & Local \\
VAD           & Silero VAD                     & Local  \\
\addlinespace[2pt]

\midrule
\rowcolor{groupgray}
\multicolumn{3}{@{}l}{\textbf{Speech Understanding}} \\
ASR           & Paraformer               & Local  \\
              % & xtalk\_zipformer\_onnx                & Local  \\
              & SenseVoice               & Local  \\
              & Qwen3-ASR-Flash               & API  \\
Environment     & Qwen3-Omni-Captioner           & Local \\
Speaker       & Wespeaker-Resnet34       & Local  \\
Paralinguistics & Qwen3-Omni-Captioner & Local  \\
                & emotion2vec  & Local  \\
\addlinespace[2pt]

\midrule
\rowcolor{groupgray}
\multicolumn{3}{@{}l}{\textbf{LLM}} \\
Dialog models &All models compatible with OpenAI endpoint           & Local/API \\
Embeddings    & All models compatible with OpenAI endpoint           & Local/API  \\
\addlinespace[2pt]

\midrule
\rowcolor{groupgray}
\multicolumn{3}{@{}l}{\textbf{Speech Generation}} \\
TTS           & IndexTTS 1.5          & Local  \\
              & IndexTTS 2        & Local  \\  
              & CosyVoice-3-Flash                & API  \\
              & GPT-SoVITS                & Local  \\
              & ElevenLabs TTS                 & API \\
              & Edge TTS                       & API \\
%Speed         & Rubberband Speed Controller    & Local  \\
\bottomrule
\end{tabular}
\end{adjustbox}
\caption{Models Integrated in X-Talk, covering four functional modules for the speech-to-speech dialogue pipeline. }
\label{tab:integrated_modles}
\end{table}

\subsection{Model-Agnostic Design}
% X-Talk is explicitly designed to be model-agnostic, avoiding assumptions about internal model architectures, execution granularity, or state representations. This is achieved by enforcing standardized interface protocols for multimodal data exchange, isolating session-specific mutable runtime states through lightweight pipeline abstractions, and coordinating heterogeneous components via an event-driven control layer rather than hard-coded timing logic. 
X-Talk is built on a model-agnostic philosophy, ensuring that the core architecture remains invariant to changes in specific model implementations, execution granularities, or internal state representations. This is achieved through three key architectural pillars:

\paragraph{Interface Protocols and Data Normalization} Capability protocols are defined as the sole interaction boundary between the managers layer and models layer, specifying data formats and modality conventions without assuming any internal model architecture or execution granularity. For example, ASR components may consume normalized audio streams (e.g., mono PCM), while TTS components emit audio in standardized formats, enabling interchangeable use of heterogeneous implementations. Optional metadata, such as suggested chunk sizes, may be exposed to facilitate efficient buffering; however, such hints remain advisory and do not constrain model behavior, preserving a stable and model-agnostic interface.

\paragraph{State Isolation through Pipeline Abstraction} Processing pipelines are implemented as lightweight abstractions whose primary role is to isolate session-specific mutable runtime states rather than encode model-specific execution logic. Immutable model parameters are shared across pipelines, while all mutable states-such as streaming buffers, and intermediate queues-are instantiated on a per-session basis. This separation ensures that concurrent sessions can safely reuse the same underlying model weights without interference, regardless of how individual models represent or manage their internal states.

\paragraph{Event-Driven Coordination with Priority Routing} Component interaction is coordinated through a centralized, event-driven control layer that routes semantic control signals (e.g., start, pause, stop, flush) based on session context and configurable priorities. Importantly, the managers layer makes no assumptions about timing, segmentation, or decoding granularity. Instead, pacing is determined entirely by individual components—such as chunk-level emission from ASR, token-level streaming from LLMs, or synthesis cadence from TTS—and outputs are forwarded as they become available. This design enables consistent coordination across heterogeneous execution models without hard-coded scheduling logic.

\subsection{Integration of New Functionalities} 
X-Talk offers great scalability and enables easy integration of new functionalities. Here we introduce two extension approaches: the first involves directly implementing the target functionality with the option to encapsulate it as a tool for the LLM to invoke; the second involves integrating new categories of models into the system to handle more complex or specialized task requirements. Together, these approaches enable individual models or services to be substituted, upgraded, or omitted without modifying the overall orchestration logic.

\paragraph{New Functionalities and New Tools} We first introduce the approach of directly inserting new functionalities. For functionalities that can be naturally integrated into the existing architecture, we can simply find out their appropriate system layer (such as a specific manager or model implementation) and then implement and integrate them at the corresponding location. During this process, new types of events can be defined as needed, and the event subscription relationships and behaviors of each manager can be adjusted. For more complex functionalities, first develop them as classes or modules. Then initialize and invoke them within the relevant manager. Additionally, the functionality can be encapsulated as a tool callable by the LLM. Simply add a new tool implementation under the agent, inherit the constraints from the interface, and define input fields, output structures, and other parameters. Subsequently, add the tool to the agent's available list to enable its invocation.

\paragraph{New Model Category} X-Talk offers great extensibility, which allows new categories of models to be easily introduced to implement additional functionality. Specifically, the following steps can be taken to integrate new categories of models. First, we need to define a new abstract base class for this model category within the interfaces layer and specify its inputs, outputs, and fundamental methods. After that, concrete model classes can be implemented based on this base class. Next, mount the model instance onto the Pipeline. Ensure that it acts as a context dependency that can be passed and supports copy operations. Following this, we need to implement a dedicated Manager (e.g., TTS\_Manager) to subscribe to relevant events on the Event Bus. This manager will invoke the model to execute business logic and publish new events via EventBus. In addition, additional event types can be registered as needed during this process. Finally, we need to register the Manager within the service's initialization list. This ensures that it starts with the session and is integrated into the unified lifecycle management and resource release system.

\section{Interaction Latency}
% \subsection{Low Latency through Event-Driven Parallel Execution}
% X-Talk adopts a session-level event-driven architecture to achieve strict end-to-end latency control. A centralized event bus asynchronously coordinates processing signals, enabling different functional modules to operate independently. The system is designed to minimize delay in the forward processing chain, which comprises ASR, LLM inference, and TTS synthesis components. Importantly, latency can be actively tuned by substituting models within this chain.

% Other functionalities-such as voice caption, speaker identification, and retrieval-triggered callbacks-are executed through parallel, non-blocking channels. This design prevents operations in one processing path from delaying those in another, ensuring that the response latency of the system is effectively bounded and predictable. 

\subsection{Asynchronous Parallel}
X-Talk adopts an event-driven architecture to achieve strict end-to-end latency control. 
A centralized event bus asynchronously coordinates processing signals, enabling different functional modules to operate independently. 
The system is engineered to minimize the critical path of the forward processing chain, which consists of ASR, LLM inference, TTS, and other modules. 
Other modules, such as audio caption, speaker identification, and retrieval-based callbacks, are executed through parallel, non-blocking channels. 
This design prevents operations in one processing path from delaying those in another, ensuring that the response latency of the system is bounded and predictable.

\subsection{Experiments}
\paragraph{Implement Details} We evaluate the end-to-end latency of X-Talk across different combinations of ASR, LLM, and TTS models, with input speech ranging from 5 seconds to 1 minute. For each component, we evaluate multiple models, including Paraformer, SenseVoice for ASR~\citep{paraformer,an2024funaudiollmvoiceunderstandinggeneration}, Qwen3-Next-80B-A3B-Instruct, Qwen3-30B-A3B, Qwen3-8B for LLM~\citep{qwen3}, and IndexTTS 1.5, IndexTTS 2, CosyVoice for TTS~\citep{deng2025indextts,zhou2025indextts2,du2024cosyvoice}. All experiments are conducted on a deployment with four NVIDIA RTX 4090 GPUs. We deploy SenseVoice using sherpa-onnx\footnote{\url{https://github.com/k2-fsa/sherpa-onnx}} and evaluate the end-to-end latency under both its original non-streaming synthesis and our pseudo-streaming implementation. All local LLMs are deployed using vLLM~\citep{kwon2023efficient}. We set the sampling temperature of the model to 0.1 to reduce randomness. Meanwhile, we set the temperatures of the Thinker and Rewriter to 0.7 to encourage broader exploration. For Qwen3-30B-A3B and Qwen3-8B, we use their quantized variants. We access Qwen3-Next-80B-A3B-Instruct and CosyVoice-v3-flash via their official APIs. For IndexTTS 1.5 and IndexTTS 2, we deploy them using vLLM\footnote{\url{https://github.com/Ksuriuri/index-tts-vllm}}. For each configuration under different speech input, we measure the latency three times and use the average value as the final result. Notably, the frontend VAD uses a 500 ms end-of-utterance silence threshold to determine the speech endpoint, and this fixed delay is not included in latency calculation there.

\paragraph{Results} In Table~\ref{tab:latency_results}, we demonstrate the end-to-end latency of X-Talk under different combinations of ASR, LLM, and TTS models. X-Talk's default configuration uses SenseVoice(Streaming), Qwen3-30B, and IndexTTS 1.5. Our system latency is primarily composed of three components: ASR full-text recognition latency, LLM first-sentence generation latency, and TTS first-sentence synthesis latency. When using Paraformer and our pseudo-streaming method with SenseVoice, the ASR full-text recognition latency becomes extremely low and is largely insensitive to the input audio length. As a result, the overall latency is mainly dominated by the LLM’s delay to generate the first sentence and the TTS’s delay to synthesize the first sentence. In this setting, the observed differences in latency are primarily driven by variations in the length of the model’s first sentence, as well as the inference speed of the LLM and TTS models. Therefore, when we switch the model to Qwen3-80B, IndexTTS2, and CosyVoice, the latency increases. Notably, for Qwen-8B, we observe that it tends to invoke tools more proactively before producing a response when presented with different speech inputs (e.g., switching emotion), which in turn increases the overall latency. In contrast, when SenseVoice is used in offline mode for speech recognition, its latency becomes sensitive to the input duration, as reflected in our results.

\begin{table}[t]
  \centering
  \small % 将字号设为 small 以节省宽度，如果仍然太宽可改为 footnotesize
  \setlength{\tabcolsep}{3pt} % 调整列间距
  \begin{tabular*}{\textwidth}{@{\extracolsep{\fill}}lllcccccccc}
    \toprule
    \multirow{3}{*}{\textbf{ASR (streamable)}} & \multirow{3}{*}{\textbf{LLM}} & \multirow{3}{*}{\textbf{TTS}} & \multicolumn{8}{c}{\textbf{Latency (ms)}} \\
    \cmidrule(lr){4-11}
    
    & & & \multicolumn{2}{c}{5s} & \multicolumn{2}{c}{10s} & \multicolumn{2}{c}{30s} & \multicolumn{2}{c}{60s} \\
    \cmidrule(lr){4-5} \cmidrule(lr){6-7} \cmidrule(lr){8-9} \cmidrule(lr){10-11}
    
    & & & CN & EN & CN & EN & CN & EN & CN & EN \\
    \midrule
    SenseVoice(streaming) & Qwen3-30B & IndexTTS 1.5 &284 & 272&346 &406 &420 &610 &449 &376 \\
    \hdashline
    \textbf{\textit{Paraformer(streaming)}} & Qwen3-30B & IndexTTS 1.5 &264 &367 & 399&488 &379 &412 &393 &382 \\
    \textbf{\textit{SenseVoice(offline)}} & Qwen3-30B & IndexTTS 1.5 &635 &617 &1015 &1054 &2175 &2326 &3798 &3671 \\
    SenseVoice(streaming) &  \textbf{\textit{Qwen3-8B}} & IndexTTS 1.5 &808 &339 &948 &625 &1039 &1111 &1075 &1068 \\
    SenseVoice(streaming) &  \textbf{\textit{Qwen3-80B(API)}} & IndexTTS 1.5 &881 &1436 &994 &1216 &1052 &1124 &1136 &1374 \\
    SenseVoice(streaming) & Qwen3-30B &  \textbf{\textit{IndexTTS 2}} &1515 &1264 &1448 &1568 &1764 &2131 &1832 &1810 \\
    SenseVoice(streaming) & Qwen3-30B &  \textbf{\textit{CosyVoice(API)}} &652 &587 &912 &1029 &763 &842 &856 &1019 \\
    \bottomrule
  \end{tabular*}
  \caption{X-Talk's latency test across different combinations of models with different input speech length. CN/EN indicates Chinese/English speech inputs. CosyVoice refers to CosyVoice-v3-flash. For each configuration, latency is measured three times and the final value is the rounded average. The first row is the baseline configuration in X-Talk, while each of the other configurations has one model varied.
  % The frontend VAD uses a 500 ms end-of-utterance silence threshold to determine the speech endpoint, and this fixed delay is not included in latency calculation there.
  } 
  \label{tab:latency_results}
\end{table}
\subsection{Sub-Second Streaming} 
%sub-second 用词待实验数据确认，如果不是低于1s则更换用词
X-Talk achieves sub-second response latency by implementing a comprehensive streaming mechanism across all system components. To minimize time-to-first-token (TTFT) and time-to-first-audio (TTFA), X-Talk enforces streaming behavior across model boundaries. Native-streaming ASR models emit partial hypotheses incrementally. For non-streaming ASR engines, we employ cumulative full-context re-inference, in which the growing audio buffer is repeatedly reprocessed and newly generated transcripts replace prior outputs, exposing a pseudo-streaming interface while preserving contextual stability.

LLM responses are emitted at token-level granularity and forwarded immediately. For TTS models without native input streaming capability, long utterances are segmented into short clauses. These segments are synthesized independently and played back sequentially, transforming batch-oriented synthesis into a low-latency, stream-like audio pipeline.

\subsection{Perceived Responsiveness}
To reduce user-perceived latency, X-Talk introduces a VAD-driven preemption loop that pauses ongoing client TTS playing immediately upon detecting user speech, followed by lightweight semantic verification. When a true barge-in is confirmed, active synthesis and inference modules are terminated; otherwise, suspended client TTS playing resume seamlessly, providing instant user feedback.

During tool-call latency, X-Talk injects short phatic utterances (e.g., "Let me check this for you...") that are rendered in real time. This strategy masks backend delays and maintains the subjective impression of continuous system activity.

\subsection{False Interruption Tolerance through Rules}
To further reduce false interruptions, X-Talk has implemented a series of rules on the backend for validation. First, the system performs an audio duration check. Specifically, X-Talk sets a default minimum audio length threshold of 0.5 seconds, and any audio shorter than this threshold is classified as a false interruption. Second, the system filters based on the ASR transcription result. If the output of ASR  is an empty string or contains only a single letter or digit, it is considered a false interruption. Additionally, X-Talk maintains a list of filler words. If the transcription result only contains words from the fillered words list, it will be considered invalid. When the backend identifies a false interruption, it notifies the frontend to resume audio playback. Through this mechanism to detect false interruptions, X-Talk achieves more natural interaction.
\section{Cross-Platform Deployment}
\subsection{Parallel Services and Concurrent Control}
X-Talk adopts an event-driven architecture. Events are published via an event bus to all subscribed handlers and executed concurrently. This architecture supports decoupling and parallel execution of core components. Event prioritization can be used to determine the processing order. Additionally, X-Talk implements multiple concurrently running service managers. For example, the TTS-Manager synthesizes speech within an independent consumer task and streams speech chunks to the output gateway without blocking LLM generation or ASR processing. This design ensures low latency and high resource utilization on the server side. Furthermore, X-Talk implements session isolation and concurrency control. The ServiceManager creates a dedicated Service instance for each WebSocket connection, which ensures sessions remain isolated and do not interfere with others. Furthermore, SessionLimiter can be used to restrict the number of concurrent sessions to reduce the load on the server.

\subsection{Communication via WebSocket} 
X-Talk uses WebSocket as the communication protocol between the frontend and the backend. WebSocket has excellent compatibility across various deployment environments and is supported by different modern browsers and application frameworks on desktop and mobile platforms. Moreover, WebSocket is well-suited for resource-constrained scenarios such as deploying lightweight clients on robots. Based on WebSocket, our backend can provide a unified service for clients of different types without requiring specific adaptations for various platforms. Additionally, compared to WebRTC, WebSocket has superior performance in network traversal. And it operates over TCP connections and ensures stable communication quality.
\section{Future Work}
In future work, we will continue to expand and enhance the capabilities of X-Talk to achieve more natural and immersive interactions and provide a greater user experience. For example, in the Speech Interaction Frontend, we will further introduce models such as acoustic echo cancellation to improve the X-Talk’s stability. And we will also introduce more types of generative models to achieve more flexible and expressive audio output.

X-Talk has shown great interactive capabilities and can be adapted to a variety of real-world application scenarios in the future. In the example from our code, the system can serve as an auxiliary role for psychological counseling, guiding users through psychological assessment questionnaires and providing a gentle interactive experience. In the future, we will continue to explore the potential applications of X-Talk in various real-world scenarios.
% \paragraph{Train Models Catered to X-Talk} 
Besides that, we plan to further improve X-Talk's interactive experience by training models that are better suited to its applications. For example, we aim to develop context-aware ASR models that integrate dialogue history and other contextual signals to reduce recognition error rates.
% For example, we plan to develop  the following models:
% \begin{itemize}
%     \item Turn detection models: These can more accurately predict speaker intentions, precisely determine turn transitions, and distinguish between pauses and speech endings.
%     \item LLMs optimized for paralinguistic information, enabling X-Talk to better understand features such as intonation, hesitation, and emotional tone.
%     \item Context-aware ASR models: By integrating dialog history and other contextual information, these models can reduce recognition error rates.
% \end{itemize}

\section{Conclusion}
% In this paper, we present X-Talk, an open-source spoken dialogue system framework tailored for real-world deployment. Built upon the modular decomposition and an event-driven architecture, X-Talk not only enables low-latency, paralinguistic-aware, full-duplex spoken interaction but also offers high modularity: individual components can be independently maintained, retrained, or optimized, and new modules can be seamlessly integrated without retraining the overall pipeline. Its performance demonstrates the potential of cascaded speech-to-speech dialogue systems through careful architectural design and the integration of various advanced speech models.

In this paper, we highlight the untapped potential of modularized spoken dialogue systems, demonstrating that such architectures can simultaneously achieve low latency and paralinguistic-aware interaction, capabilities often regarded as the primary strengths of end-to-end “omni” models. We present X-Talk, a highly optimized cascaded dialogue system that delivers superior performance in terms of latency, barge-in, complex audio understanding, and information retrieval—areas that remain challenging for most existing open-source end-to-end systems. Moreover, X-Talk is inherently extensible, enabling seamless integration of new modules and stronger models as they emerge, thereby lowering the barrier for community contribution and continuous evolution. We believe that X-Talk provides a practical and scalable foundation for advancing LLM-driven speech-to-speech applications.

\section*{Acknowledgment}

We express sincere gratitude towards Fangjun Kuang, one of contributors towards sherpa-onnx, for helping us getting speech recognition batch inference work properly.

% \applefootnote{ \textcolor{textgray}{\sffamily xxxxx.}}

\bibliographystyle{plainnat}
\bibliography{biblio}

\clearpage
\appendix
% ignore

\end{document}